# SRECG: ECG Signal Super-resolution Framework for Portable/Wearable Devices in Cardiac Arrhythmias Classification

Tsai-Min Chen, Yuan-Hong Tsai, Huan-Hsin Tseng, Kai-Chun Liu, Jhih-Yu Chen, Chih-Han Huang, Guo-Yuan Li, Chun-Yen Shen, and Yu Tsao*

***Abstract*—A combination of cloud-based deep learning (DL) algorithms with portable/wearable (P/W) devices has been developed as a smart heath care system to support automatic cardiac arrhythmias (CAs) classification using electrocardiography (ECG). However, long-term and continuous ECG monitoring is challenging because of limitations of batteries and transmission bandwidth of P/W devices while incorporated with consumer electronics (CE). A feasible approach to address this challenge is to decrease sampling rates. However, low sampling rates lead to low-resolution signals that hinder the CAs classification performance. In this study, we propose a DL-based ECG signal super-resolution framework (called SRECG) to enhance low-resolution ECG signals by jointly considering the accuracies when applied to the DL-based high-resolution multiclass classifier (HMC) of CAs. In our experiments, we downsampled the ECG signals from the CPSC2018 dataset and evaluated their HMC accuracies with and without the SRECG. Experimental results show that SRECG can well improve the HMC accuracies as compared to traditional interpolation methods. Moreover, approximately half of the CAs classification accuracies of HMC were maintained within the enhanced ECG signals by SRECG. The promising results confirm that SRECG can be suitably used to enhance low-resolution ECG signals from P/W devices with CE to improve their cloud-based HMC performances.**

## I. INTRODUCTION

Consumer technologies have changed the world, especially in healthcare, where traditional medicine has been pushed to transform into telemedicine, connected-health, e-health, mobile-health, and smart health [1]. Through the growth of portable/wearable (P/W) devices, smart healthcare systems enable medical doctors to not only collect patient data without worrying about the physical distance in between, but also to implement instant treatment remotely [2]. Moreover, P/W devices make it possible to continuously monitor the health conditions of a patient and collect relevant data. This data is helpful in alerting medical professionals to critical conditions and implement medical actions and is valuable for providing new medical insights for further research.

Cardiac arrhythmias (CAs) are harbingers of cardiovascular diseases that are life-threatening to human beings [3]. A typical assessment tool relies mainly on electrocardiography (ECG) signals that record the electrical activity of the heart. ECG is a noninvasive and inexpensive method that is clinically applied to monitor heart functionality [4-6]. The diagnosis of CAs is based on wave-like features, such as the P wave, QRS complex, and T wave, of ECG signals. Medical doctors can manually examine the ECG features to evaluate the occurrence of CAs. The entire examination process for a long-term ECG record is time consuming and tedious. Therefore, it is essential to develop an automatic CAs classification system to support clinical diagnosis.

Several studies have developed cloud-based CAs monitoring systems that combine cloud services and advanced deep learning (DL) models for smart healthcare systems. These CAs monitoring systems collect ECG using P/W devices and convey heavy processing to cloud DL engines for further CAs classification [7, 8]. For a P/W device incorporated with consumer electronics (CE), long-term monitoring of CAs requires continuous sampling and interpretation of user data, which may lead to unacceptable power consumption. Factors such as the monitoring source, sampling frequency, electronic components, and transmission protocols affect the power efficiency of a P/W device. Reducing the monitoring sources and sampling frequency are two compromises for reducing power consumption. The latter has been extensively studied because at high sampling

Manuscript received September xx; revised xx; accepted xx. Date of publication xx; date of current version xx. This work was supported in part by grants from the National Science and Technology Council, Taiwan, under Grant 111-2221-E-001-016-MY3, and Academia Sinica, under Grant AS-GC-111-M01.
*(Corresponding author: Yu Tsao, e-mail: yu.tsao@citi.sinica.edu.tw).*

T.-M. Chen, and Y. Tsao are with the Graduate Program of Data Science, National Taiwan University and Academia Sinica, Taipei, Taiwan; Research Center for Information Technology Innovation, Academia Sinica, Taipei, Taiwan.

Y.-H. Tsai, and G.-Y. Li are with the Program of Taiwan AI Academy, Taiwan Artificial Intelligence Academy Foundation, New Taipei, Taiwan; Technology Development Center, Artificial Intelligence Foundation, New Taipei, Taiwan.

H.-H. Tseng, and K.-C. Liu are with the Research Center for Information Technology Innovation, Academia Sinica, Taipei, Taiwan.

J.-Y. Chen is with the Graduate Institute of Biomedical Electronics and Bioinformatics, National Taiwan University, Taipei, Taiwan.

C.-H. Huang is with the Institute of Biomedical Sciences, Academia Sinica, Taipei, Taiwan.

C.-Y. Shen is with the Graduate Program of Data Science, National Taiwan University and Academia Sinica, Taipei, Taiwan; Department of Mathematics, National Taiwan University, Taipei, Taiwan.

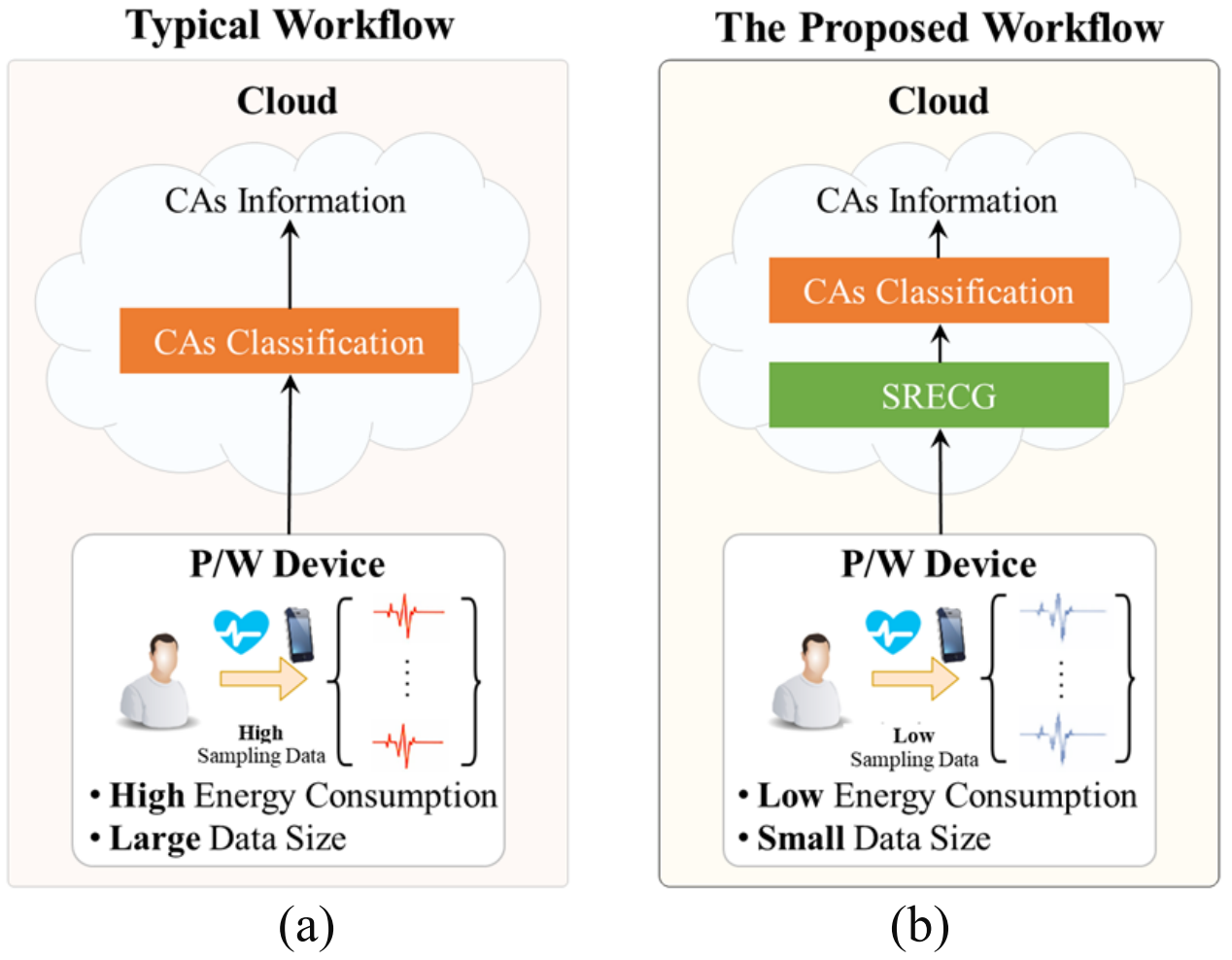

**Fig. 1. Comparison of DL Workflows of Smart Healthcare System for a P/W Device.** (a) Typical workflow. (b) The proposed workflow based on the proposed SRECG.

frequencies, up to 90% of power consumption of health monitoring systems is due to data sampling [9-11].

Various power efficiency strategies have been applied to automatic CAs classification systems to enable long-term health monitoring services [10-12]. Maji et al. proposed an adaptive power management approach to classify critical and non-critical signals before further DL-based ECG classification [13]. Hierarchical classification models can efficiently reduce power consumption [13]. Furthermore, they employed typical data augmentation approaches, including SMOTE and BIRCH [14, 15], to improve performance in terms of power consumption [13]. Biagetti et al. proposed a low-cost wireless system to transmit the acquired surface electromyographic and ECG signals through a 2.4 GHz radio link based on the IEEE 802.15.4 physical layer [16]. This system can be combined with different mobile and wearable devices to support low-cost and low-power health-monitoring systems [16]. Raj proposed a tuned twin support vector machine (SVM) classifier using particle swarm optimization to classify 16 categories of ECG signals on benchmarked Physionet data, which was four times faster than the standard SVM and showed enhanced performance when implemented on the Internet of Things platform [17].

Despite these strategies, the reduction in monitoring sources is still a typical and direct approach to saving energy, for example, reducing the recording of ECG sources from a full 12-lead ECG signal to a single lead [18, 19]. In our previous studies, we provided an empirical discussion over a 12-lead source to select the most suitable single-lead ECG source for DL-based CAs classification [20]. Another approach is to decrease the sampling rate of the P/W device, which is an efficient approach for reducing power consumption. However, in general, a lower sampling frequency is accompanied by a reduction in information, which is, low resolution. Low-resolution signals can cause degradation of CAs classification. For example, several important ECG characteristics are degraded in low-frequency signals, which may decrease the ability of the monitoring system to detect CAs automatically. Although there are studies regarding the sampling frequency of ECG, most are related to the association of heart rate variability, which is widely used as a noninvasive marker of the autonomic nervous system [21, 22].

A few existing studies have aimed to develop low-resolution automatic CAs classification. Sidek et al. proposed interpolation-based signal enhancement approaches, involving piecewise cubic Hermite interpolation and piecewise cubic spline interpolation, to increase the quality of signals and classification accuracy at a lower sampling frequency [23]. The CAs detection accuracy using the interpolation-based enhancement approach could reach ~99%, whereas without signal enhancement, it only reaches ~97% [23]. Mathews et al. proposed a DL-based CAs classification framework using a Restricted Boltzmann Machine and deep belief networks to improve the classification abilities at lower sampling rates, where a range of sampling frequencies from 72 Hz to 360 Hz was explored [24]. The experimental results validated the DL approach for low-resolution CAs classification [24]. However, these previous studies did not investigate automatic CAs classification at very low sampling rates, while several studies have developed P/W-based monitoring systems with an extremely low sampling frequency to support other long-term healthcare services, such as daily activity recognition and fall detection [10, 25].

To alleviate the loss of information at very low sampling frequencies, image super-resolution (SR) may serve as a viable solution to enhance low-resolution ECG signals for automatic CAs classification. Image SR has recently become a popular and well-studied area of interest [26-28]. The idea is to reconstruct the target information from the low-resolution data. The primary goal of image SR is to recover data from compressed or low-quality signals to facilitate better signal resolution for subsequent tasks. In particular, image SR has long been an important subject of image processing techniques in computer vision that aim to recover high-resolution images from low-resolution images, which in turn is a challenging task owing to its mathematically ill-posed nature [29-32]. The rapid development of DL techniques in recent years has also driven researchers to tackle image SR tasks. In general, DL-based SR algorithms have three different components: (1) network architectures, such as convolutional layers, residual connections, recursive layers, and up-sampling layers [33-35]; (2) design of loss functions, such as pixel, perceptual, or adversarial losses [31, 36, 37]; and (3) learning principles and strategies [38-41]. Some SR algorithms have reached the current state-of-the-art performance [42, 43], proving useful for improving other vision tasks [29-32]. Task-driven SR can significantly enhance the accuracy of an object detector on low-resolution images, thus creating a positive impact on vision recognition [30]. Similarly, in signal processing, audio SR refers to the task of increasing the sampling rate for a given low-resolution (i.e., low-sampling rate) audio. Motivated by recent advances in learning-based algorithms for speech recognition [44-46], music generation [47, 48], and other areas [49], significant progress has been made in audio SR via the introduction of DL. The key advantage here is that one is allowed to directly model raw signals in the time domain [47, 50-52], and effectively capture long-term dependencies [53].

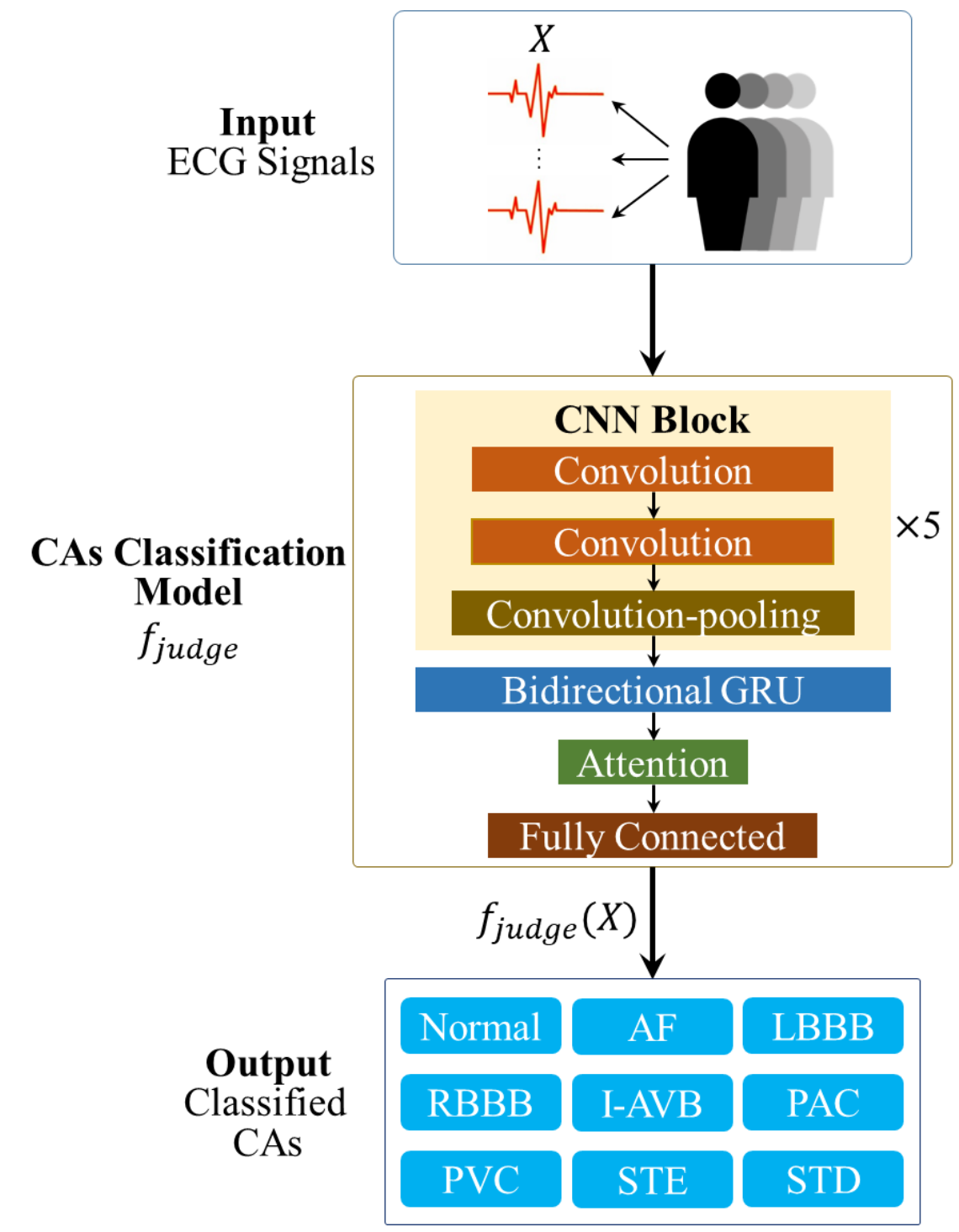

**Fig. 2. Architecture of CPSC2018 benchmarked DL network** (denoted by the function $f_{judge}$). Layers and blocks are specified in rectangle boxes; "×5" indicates that five CNN blocks are tandem connected before being connected to the bidirectional GRU layer. The output layer at the bottom contains the probabilities predicted by the model for each of the nine types of the CAs classification. The type with the highest probability is the type predicted by the model for the input of an ECG signal.

Although SR has had a leaping success in human vision tasks, to the best of our knowledge, no prior studies have attempted to apply the SR technique to ECG signals for automatic CAs classification. In this paper, we proposed and investigated a novel DL-based ECG signal SR framework (SRECG) as a processor. This framework is expected to help automatic CAs classification tackle the challenges of low-resolution ECG for P/W devices and achieve better accuracy. The proposed SRECG enhances low-resolution ECG signals with composite training loss in the cloud and, in turn, yields better utilization for the DL-based high-resolution multiclass classifier (HMC) of CAs. As shown in Fig. 1, compared to the typical workflow of a smart healthcare system, when employing SRECG as a computing processor, the ECG signals can be sampled at extremely low frequencies to reduce both energy consumption and data size for data transmission. The processor and HMC are both in the cloud; therefore, their computing power consumption is separated from that of a P/W device. Furthermore, we comprehensively analyzed and explored the impact of sampling frequencies on the DL-based CAs classification of 12 different lead sources of the ECG signals. It would be beneficial to develop reliable DL-based CAs classification models for smart healthcare systems using P/W devices.

The main contribution of this work is threefold: (1) it is the first work that investigates the ECG signals in the CAs classifier among different leads and sampling frequencies, which clearly shows the information loss with resolution reduction of ECG signals sampled from 500 to 1 Hz; (2) our SRECG framework enhances the CAs classification accuracies of HMC when using low-resolution ECG signals from a P/W device at low sampling frequency to resolve its energy consumption and transmission issue with CE; (3) training SRECG with composite loss can recover CAs classification accuracies of HMC better than using only single loss. We believe the proposed SRECG framework could be readily adapted for commercial applications in ECG diagnostics, where data storage is an important consideration.

## II. Materials and Methods

In this study, we attempted to enhance the ECG signals from 12 different lead sources at low sampling frequencies. Therefore, the open-source training dataset, benchmarked DL model, and categorical metric for performance evaluation are all from CPSC2018, which is a worldwide automatic ECG CAs classification competition [54].

### *A. CPSC2018 Open-source Dataset*

Detailed information on the CPSC2018 ECG database can be found in [54] by Liu et al. In comparison to the source of the publicly available ECG database with only 2-lead ECG signals of 48 recordings for most previous CAs prediction studies, such as the MIT-BIH Arrhythmia Database [55, 56], it contains 12-lead ECG signals of 6,877 recordings with different durations from seconds to minutes and a label of nine categories, including a normal type and eight abnormal CAs: AF, left bundle branch block (LBBB), right bundle branch block (RBBB), first-degree atrioventricular block (I-AVB), premature atrial contraction (PAC), premature ventricular contraction (PVC), ST-segment elevation (STE), and ST-segment depression (STD). Of the 6,877 recordings, 476 received more than one CA-type label. To simulate the measurements of ECG signals from different lead sources at a low sampling frequency, the CPSC2018 open-source ECG dataset with a single CAs label (6,401 out of 6,877 recordings) was downsampled from 500 Hz to 250, 125, 100, 50, 25, 10, 5, 2, and 1 Hz, and the signals of different lead sources were extracted from 12-lead signals separately.

The down-sampling approach was employed to obtain low-resolution signals of different lead sources from high-resolution signals to explore the effects of different resolutions on the CAs classification performance. The low-resolution signals are gathered by taking an integer factor $n$, which refers to the skipping number of sequential units of time to collect the signal. In this study, $n = 1, 3, 4, 9, 19, 49, 99, 249,$ and $499$ is applied to the CPSC2018 dataset to obtain sampling frequencies: 250, 125, 100, 50, 25, 10, 5, 2, and 1 Hz. To compensate for the unequal length of recordings, we simply applied zero padding before any recording that was shorter than the required length of maximum time of the whole CPSC2018 dataset. There was no preprocessing for the recordings.

### *B. CPSC2018 Open-source Benchmarked DL Network*

Our previous model architecture, winning CPSC2018, was selected as the reference model to create our HMCs and test the CAs classification performances at different sampling frequencies [20]. It was built on a combined architecture of five CNN blocks, followed by a bidirectional gated recurrent unit (GRU), an attention layer [57, 58], and a fully connected layer (see Fig. 2). In our architecture, each CNN block contained two convolution layers that were followed by a convolution-pooling layer, which is a convolution-based pooling layer, to reduce the number of parameters and computation in the network and to control overfitting [59]. All convolution layers share the same kernel number of 12 and kernel size of three, except for the five convolution-pooling layers with kernel sizes of 24, 24, 24, 24, and 48, sequentially. Furthermore, between these CNN blocks and other independent layers, including the one between the last CNN block and the bidirectional GRU layer, 20% of their connections were randomly dropped. Batch normalization was also used to adjust and rescale the results from the attention layer, which is a special mechanism proposed to generate importance-weighting vectors [60]. LeakyReLU activation functions were used in each layer other than the fully connected layer, where a sigmoid activation function was used [61]. This architecture was built using the Keras package supported by TensorFlow in GPUs [62, 63]. The model has 28,035 trainable parameters to ensure it is compact enough to explore the statistical relations using limited computing resources [20].

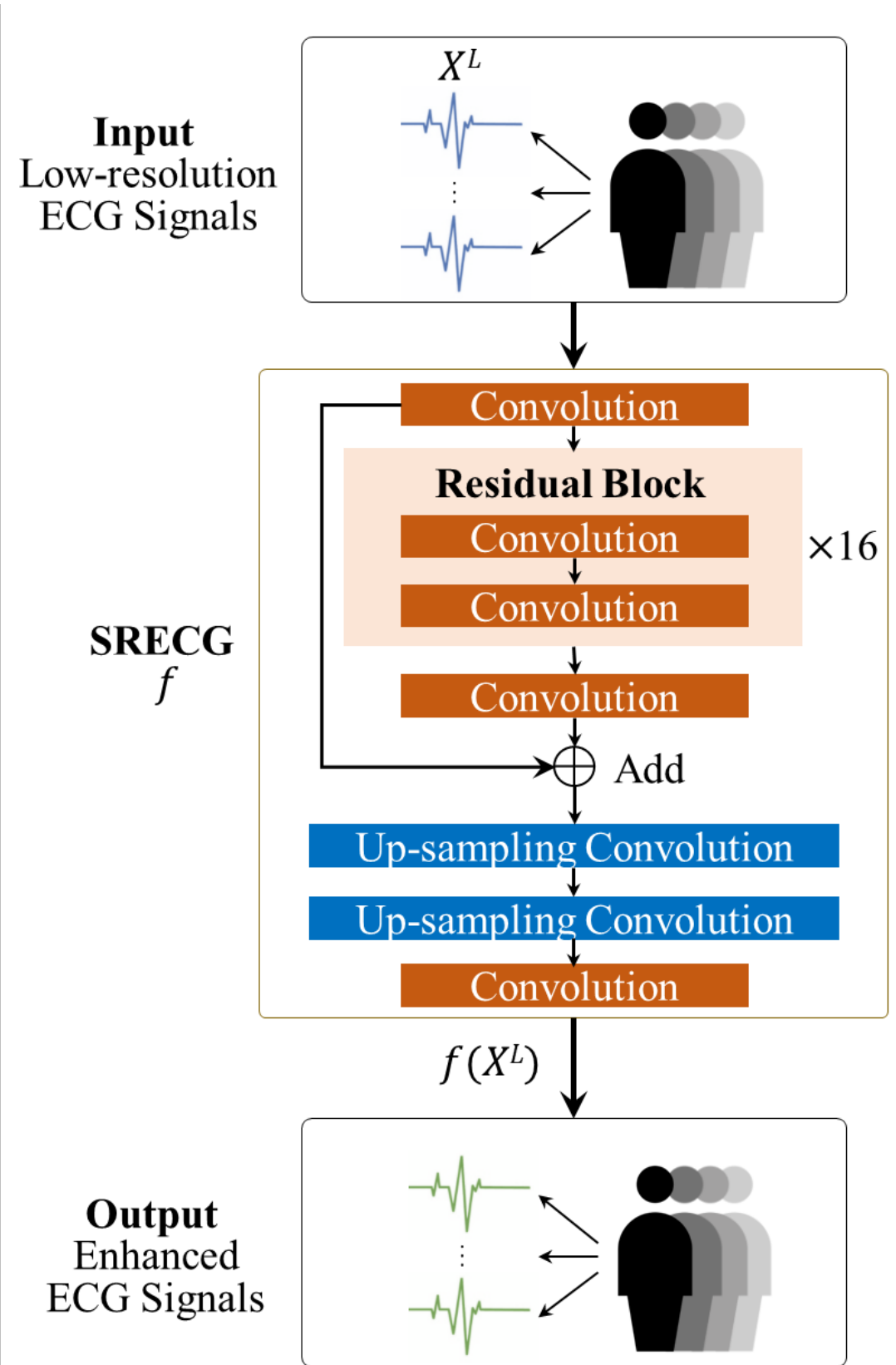

**Fig. 3. Architecture of SRECG** (denoted by the function $f$). $X^L$ is the low-resolution ECG signals input, and $f(X^L)$ is the model output as the enhanced ECG signals. Layers and blocks are specified by the rectangles; "×16" indicates that 16 residual blocks are tandem connected before joining the convolution layer.

### C. Performance evaluation

1.) F1-score

F1-score is an evaluation metric that considers both Precision and Recall by:

$$F1 = 2 \times \frac{Precision \times Recall}{Precision + Recall}, \tag{1}$$

$$Precision = \frac{TP}{TP+FP}, \tag{2}$$

$$Recall = \frac{TP}{TP+FN}, \tag{3}$$

where true positive (TP), false positive (FP), true negative (TN), and false negative (FN) are defined by

$$TP_j = \left|\left\{(x_i,\ y_i) \middle| y_i = f_j(x_i) = 1, 1 \le i \le N\right\}\right|, \tag{4}$$

$$FP_j = \left|\left\{(x_i,\ y_i) \middle| y_i = 0 \ and \ f_j(x_i) = 1, 1 \le i \le N\right\}\right|, \tag{5}$$

$$TN_j = \left|\left\{(x_i,\ y_i) \middle| y_i = f_j(x_i) = 0, 1 \le i \le N\right\}\right|, \tag{6}$$

$$FN_j = \left|\left\{(x_i,\ y_i) \middle| y_i = 1 \ and \ f_j(x_i) = 0, 1 \le i \le N\right\}\right|, \tag{7}$$

where $(x_i,\ y_i)$ is the input and label of the ith sample out of N samples, along with the prediction from a $C$-categorical classifier $f = (f_1, \dots f_j, \dots, f_C) \in \{0,1\}^C$; each categorical prediction $f_j$ is binary-valued, and $|\cdot|$ measures the cardinality of a set. The overall F1-score can be derived by averaging the F1-scores of all categories.

2.) Predictive Power Recovery

To quantify the recovery of the CAs classification accuracies of the HMC by its high-resolution ECG input with the enhanced low-resolution ECG input, a new evaluation metric called predictive power recovery (PPR) is defined as follows:

$$\mathrm{PPR} = \frac{\mathrm{E_R - C_N}}{\mathrm{C_P - C_N}}, \tag{8}$$

where $C_P$ and $C_N$ are the F1-scores of the DL models as positive and negative controls, respectively, in the CAs classification trained at high-resolution and low-resolution ECG signals, respectively. $E_R$ is the F1-score of the HMC with an enhanced low-resolution ECG input. Ideally, the relation $0 \le \mathrm{C_N} \le \mathrm{E_R} < \mathrm{C_P} \le 1$ is expected to hold, such that $0 \le \mathrm{PPR} < 1$. However, cases exist in which the relation fails (see Fig. 7), and the corresponding discussions can be found in Experimental Results.

3.) 10-fold Cross-validation Procedure of Machine Learning

The data were randomly divided into 10 equal parts to set up an 8-1-1 training, validation, and test scheme for machine learning. Under such data splitting, the model was trained for 100 epochs (which refers to one complete cycle feeding training dataset) to generate 100 models (one model at each epoch), and the model with the best performance on the validation set was selected as the best model for this training process. Subsequently, the best model was used to compute the F1-score on the test set. The procedure was repeated 10 times to complete a 10-fold training; thus, the 10 best models were selected for each fold. The median F1-score for the overall and each CAs label, including the normal type, for the 10 test sets was calculated using the Scikit-learn package [64]. The training process was implemented using the ADAM optimizer [65].

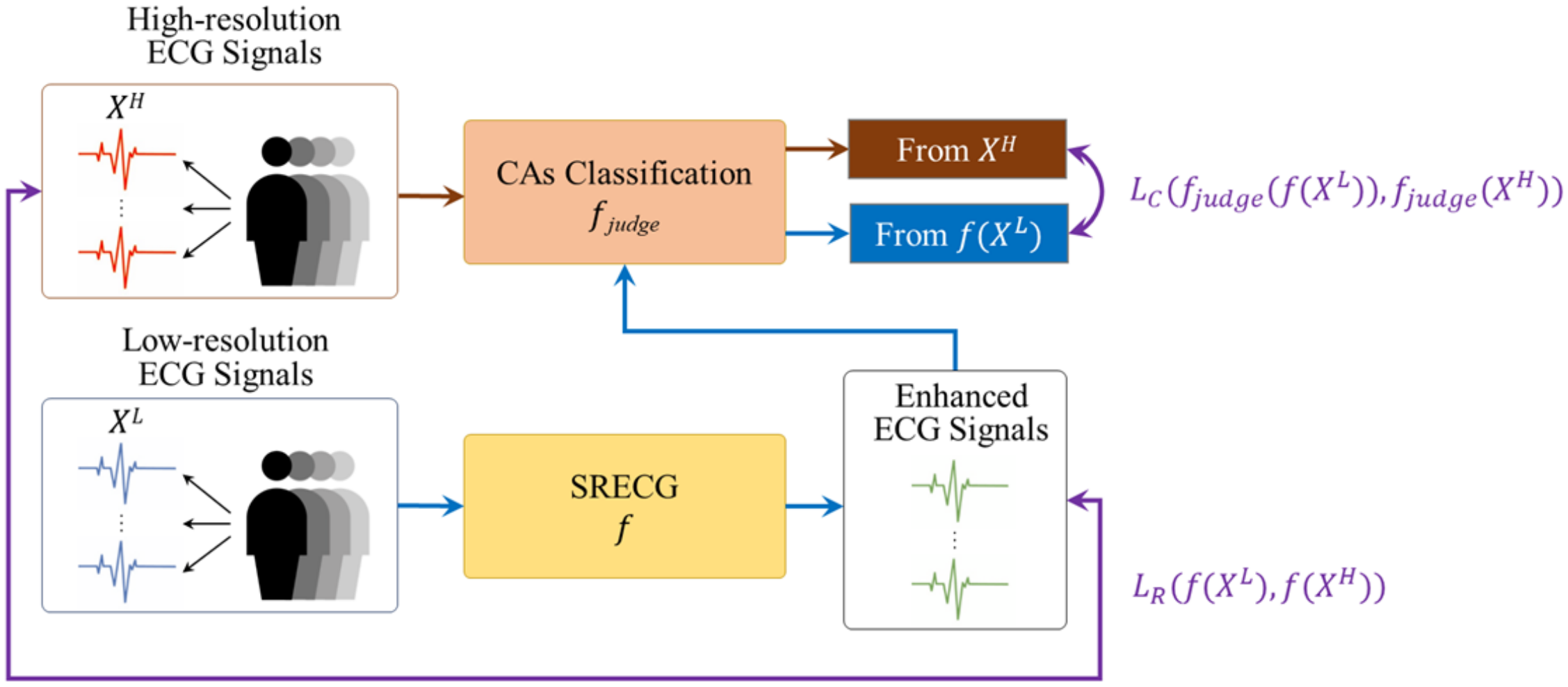

**Fig. 4. Flowchart of the proposed SRECG for enhancing low-resolution ECG data.** The SRECG $f$ receiving the low-resolution ECG data $X^L$ as input generates enhanced ECG prediction $f(X^L)$ as output by considering the joint loss $L_J$ combining the regression loss $L_R$ and the CAs classification loss $L_C$ in Equation 1. The detailed structures of DL-based CAs classification model and SRECG are depicted in Figs. 2 and 3, respectively.

### D. *Traditional Spline Interpolation for Up-sampling*

Spline interpolation is a useful technique for enhancing low-resolution ECG signals from low sampling frequencies under optimal CAs classification. Originally, spline was a term for curves bent to pass through a number of existing data points.

Spline interpolation is realized by a special type of interpolation that fits many piecewise low-degree polynomials between local data pairs, for example, fitting two polynomials between each of the pairs of three points, instead of fitting a single degree-three polynomial to all of them. Therefore, the oscillation at the edges of an interval can be reduced using low-degree polynomials for interpolation. In the numerical analysis, the spline interpolation curves are defined by a set of polynomials $\{q_i\}$ with $y = q_i(x)$ satisfying two data pairs $(x_{i-1}, y_{i-1})$ and $(x_i, y_i)$, and $i = 1, 2, \ldots, n$. There are $n$ polynomials and $n + 1$ knots in this case: the first polynomial ends at $(x_0, y_0)$, and the last polynomial ends at $(x_n, y_n)$.

### E. *Proposed SRECG*

Instead of using traditional spline interpolations to enhance the low-resolution ECG signals, we propose an SR-based method called SRECG, as shown in Fig. 3. The SRECG is modified from a well-known architecture, SRResNet [66] to have one convolution layer and 16 residual blocks [67], each of which contains two convolution layers with residual connections, followed by four convolution layers for both up-sampling and signal reconstruction. All convolution layers share the same kernel size of 16 and kernel number of 64, except for the last two up-sampling convolution layers with kernel numbers of 320 (for five times up-sampling) and 128 (for two times up-sampling), respectively. In the residual blocks, batch normalizations were also used to adjust and rescale the input from the two convolution layers (downsampling layers were not applied). LeakyReLU activation functions were applied between convolution layers before they were connected to the residual blocks and within up-sampling. Finally, batch normalization is used in the convolution layer following the residual blocks. This architecture was built using the Keras package supported by TensorFlow in GPUs [62, 63].

In Fig. 4, the low-resolution ECG data are fed into this architecture for training under the same 10-fold cross-validation procedure. Throughout the training process, the model attempts to reconstruct an enhanced ECG from a low-resolution signal. In contrast to the conventional SR task, which merely increases the resolution of an imaging system by minimizing the signal mean square error (MSE) difference, we intended to combine the CAs classification results of another DL-based judge model, which is, HMC, for assistance. Considering this goal, a special loss function was designed to simultaneously optimize the ECG signal reconstruction and predictability for CAs classification, wherein the final enhanced output is subsequently fed into the HMC for CAs judgment. Notably, the HMC was independently pre-trained with high-resolution ECG signals using the same 10-fold cross-validation scheme. The customized joining loss $L_J$ for the SRECG is defined as follows:

$$L_J(\hat{y}, y, \hat{z}, z) = \gamma\, L_R(\hat{y}, y) + (1 - \gamma)\, L_C(\hat{z}, z), \gamma \in [0,1], \quad (9)$$

where $L_R$ denotes the MSE regression loss between the high-resolution ECG signals $y$ and the enhanced ones $\hat{y}$:

$$L_R(\hat{y}, y) = \frac{1}{M} \sum_{i=1}^{M} (\hat{y}_i - y_i)^2, \quad (10)$$

where $M$ denotes the total number of maximum time steps of $y$

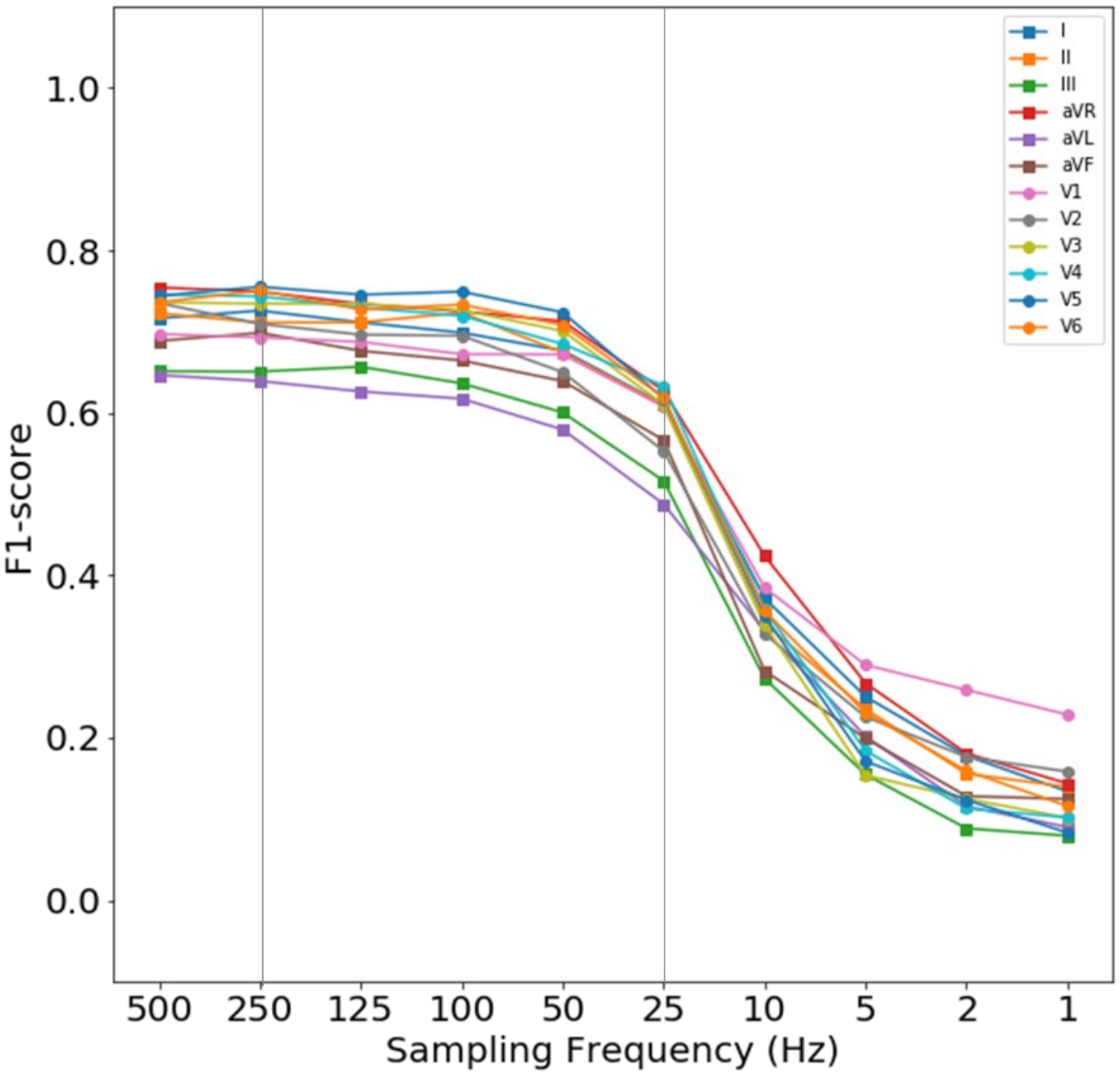

**Fig. 5. Predictive power changes of DL-based multiclass CAs classifiers trained at downsampled ECG data.** 12 lines with different colors represent the median overall F1-scores on the 10-fold tests of the CAs classification model in Fig. 2, which are trained at ECG data of different lead sources, sampled at different frequencies (500, 250, 125, 100, 50, 25, 10, 5, 2, 1 Hz).

and $\hat{y}$. The categorical cross-entropy (CCE) loss $L_C$ measures the difference between the true labels of HMC $z = (z_1, \dots, z_C) \in [0,1]^C$ and that of predictions $\hat{z} = (\hat{z}_1, \dots, \hat{z}_C) \in (0,1]^C$ by

$$L_C(\hat{z}, z) = -\sum_{k=1}^{C} z_k \log(\hat{z}_k), \tag{11}$$

In our model, we let $z = f_{judge}(y)$ and $\hat{z} = f_{judge}(\hat{y})$ serve as the CAs probabilities using a (fixed) HMC $f_{judge}$ such that the SR loss

$$L_J(\hat{y}, y) = L_J(\hat{y}, y, f_{judge}(\hat{y}), f_{judge}(y)), \tag{12}$$

is determined once the enhanced ECG signals $\hat{y}$ are generated. It is also understood that all the above losses are averaged over the sample numbers for normalization during the actual implementation.

## III. Experimental Results

### *A. Impacts of sampling frequency on DL-based CAs classification*

A range of sampling frequencies from 500 to 1 Hz of ECG signals from 12 different lead sources was tested to explore the impact of sampling frequency on the DL-based multiclass CAs classifier in Fig. 2 under the 10-fold cross-validation procedure (see Fig. 5). In our preliminary results of ECG among different leads and sampling frequencies, it was observed that the 25 Hz sampling frequency lost most of the information for the overall CAs classification, which was in contrast to the case of the 250 Hz sampling frequency that preserved the most information. Based on this observation, we applied the traditional spline interpolation method and the proposed SRECG model to enhance 25 Hz input to improve its F1-score of HMC, trained at 250 Hz.

### *B. Effects of traditional spline interpolation methods*

Four representative traditional spline interpolation methods were demonstrated for 1-dimensional signals: spline interpolation using polynomials of degree zero ($I_0$), degree one ($I_1$), degree two ($I_2$), and degree three ($I_3$). Table I shows the comparison of median overall F1-scores of CAs classification from the $C_N$ (trained at 25 Hz), $C_P$ (trained at 250 Hz), and HMC (trained at 250 Hz ECG) with spline-interpolated input from 25 Hz. Under different ECG lead sources, we observed that none of the interpolations F1-scores surpassed $C_N$. Figure 6 shows the signal reconstruction performance. The ECG waveforms of a lead II heartbeat demonstrated that $I_1$ had better reconstruction ability, resulting in the lowest MSE of 1.11% and the highest Pearson correlation coefficient (R) of 72.12%, in contrast to $I_0$ (MSE:3.71%; R:24.55%), $I_2$ (MSE:1.26%; R:71.74%), and $I_3$ (MSE:1.32%; R:70.73%).

From Table I and Fig. 6, we derived the observation that although traditional spline interpolations obtained good ECG signal reconstructions, CAs classifications (via F1-scores) were

TABLE I

MEDIAN OVERALL F1-SCORES (%) OF HMCS TRAINED AT 250 HZ WITH DIFFERENT LEADS AND SPLINE INTERPOLATION DEGREES

| Leads / degrees | $C_N$ | $I_0$ | $I_1$ | $I_2$ | $I_3$ | $C_P$ |
|---|---|---|---|---|---|---|
| *I* | **61** | 56 | 47 | 43 | 39 | 73 |
| *II* | **61** | 57 | 47 | 29 | 24 | 71 |
| *III* | **52** | 50 | 45 | 36 | 31 | 65 |
| *aVR* | **63** | 56 | 41 | 38 | 36 | 75 |
| *aVL* | **49** | **49** | 42 | 38 | 36 | 64 |
| *aVF* | **57** | 56 | 52 | 35 | 29 | 70 |
| *V1* | **61** | 59 | 50 | 45 | 42 | 69 |
| *V2* | **55** | 34 | 32 | 23 | 18 | 71 |
| *V3* | **61** | 33 | 22 | 18 | 16 | 73 |
| *V4* | **63** | 39 | 20 | 13 | 12 | 74 |
| *V5* | **62** | 45 | 18 | 14 | 12 | 76 |
| *V6* | **62** | 53 | 31 | 26 | 22 | 75 |

The bold numbers indicate the highest scores close to their $C_P$.
$C_N$: negative control (trained at 25 Hz), $I_0$: degree zero spline interpolation, $I_1$: degree one spline interpolation, $I_2$: degree two spline interpolation, $I_3$: degree three spline interpolation, $C_P$: positive control (trained at 250 Hz).

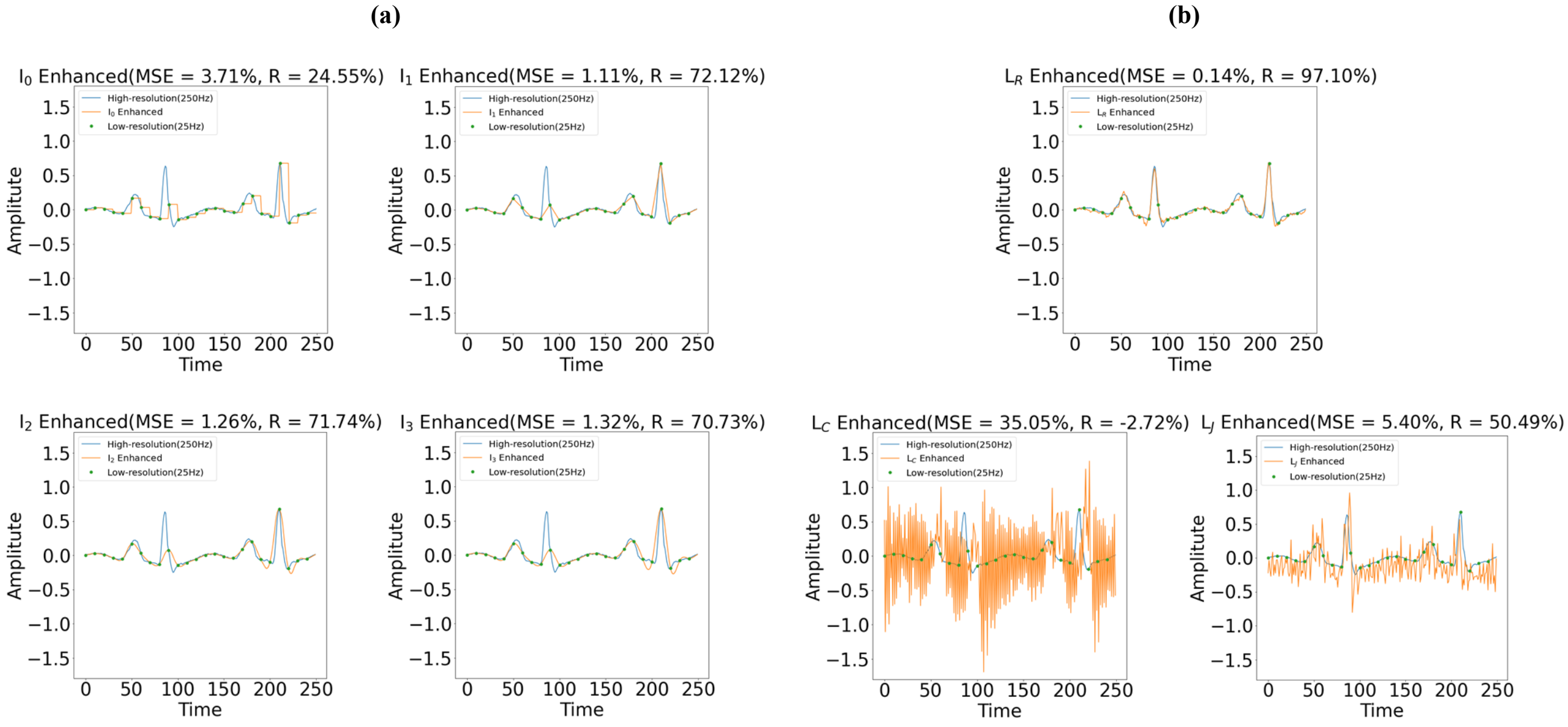

**Fig. 6. Enhanced ECG signals visualization of different spline interpolation degrees and SRECG losses.** Representative ECG waveforms of a lead II heartbeat of 1-second lapse with (a) four spline interpolation degrees: $I_0, I_1, I_2, I_3$, and (b) three losses: $L_R, L_C, L_J$. The horizontal axis has the unit of time and the vertical axis indicates the ECG amplitude. Their MSEs and Rs of one second duration are also indicated in the subtitles to show the corresponding reconstruction quality.

better preserved by the SRECG model (see $C_N$ in Table II), which indicates that traditional spline interpolation does not recover sufficient CAs information during ECG signal enhancement.

### C. *Joint effects of two different training losses for SRECG: reconstruction and classification*

The composite loss function of the SRECG in Equation 9 was designed to simultaneously optimize two distinct objectives: CCE for CAs classification and MSE for signal reconstruction. Therefore, the ratio of the two losses can be weighted by a constant $\gamma \in [0,1]$. We demonstrate three representative loss conditions in Equation 9: pure regression loss ($L_R$), pure classification loss ($L_C$), and half-mixture joint loss ($L_J$) by $\gamma$ =1, 0, and 1/2, respectively.

Table II shows the comparison of the median overall F1-scores of $C_N$ (trained at 25 Hz), $C_P$ (trained at 250 Hz), and the HMC, trained at 250 Hz ECG, with SRECG enhanced input from 25 Hz by various losses.

TABLE II
MEDIAN OVERALL F1-SCORES (%) OF HMCS TRAINED AT 250 HZ WITH DIFFERENT LEADS AND SRECG LOSSES

| **Leads / losses** | $C_N$ | $L_R$ | $L_C$ | $L_J$ | $C_P$ |
|---|---|---|---|---|---|
| *I* | 61 | 60 | 65 | **66** | 73 |
| *II* | 61 | 64 | 64 | **66** | 71 |
| *III* | 52 | 51 | 56 | **58** | 65 |
| *aVR* | 63 | 68 | **71** | **71** | 75 |
| *aVL* | 49 | 48 | 55 | **56** | 64 |
| *aVF* | 57 | 59 | 61 | **62** | 70 |
| *V1* | 61 | 62 | **65** | 64 | 69 |
| *V2* | 55 | 62 | **64** | 61 | 71 |
| *V3* | 61 | 65 | **67** | 65 | 73 |
| *V4* | 63 | 65 | 65 | **68** | 74 |
| *V5* | 62 | 65 | **70** | **70** | 76 |
| *V6* | 62 | 57 | **69** | **69** | 75 |

The bold numbers indicate the highest scores close to their $C_P$.
$C_N$: negative control (trained at 25 Hz), $L_R$: SRECG by pure regression loss, $L_C$: SRECG by pure classification loss $L_J$: SRECG by half-mixture joint loss, $C_P$: positive control (trained at 250 Hz).

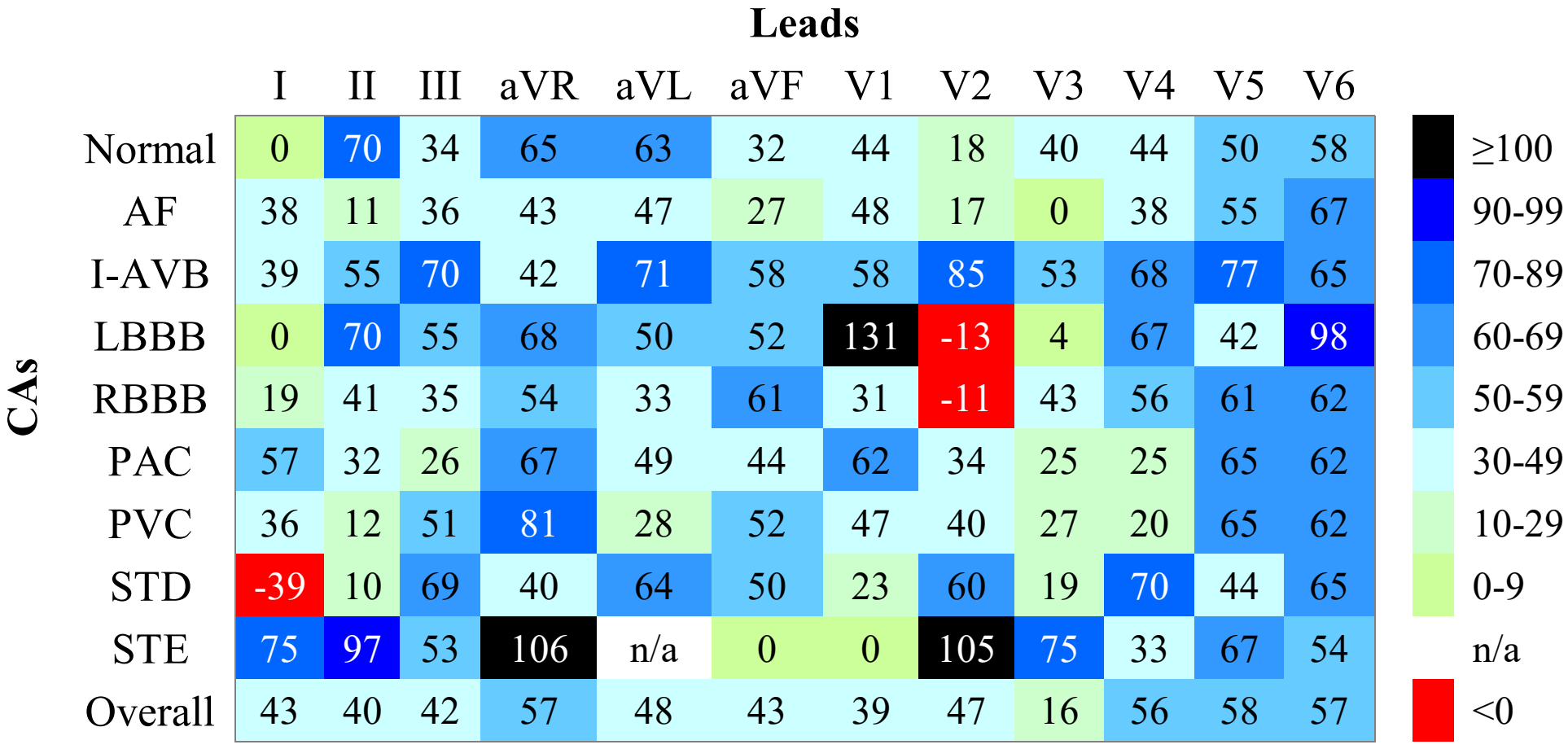

**Fig. 7. Median PPRs (%) of HMCs trained at 250 Hz with SRECG by $L_J$.** Blue labels and green labels indicate relatively high and low median PPR values, respectively. Black labels indicate median PPR values greater than or equal to 100%. Red labels indicate negative median PPR values (<0%). The white label indicates the median PPR value that is not applicable (n/a) because its $C_P$, $C_N$, and $E_R$ are all zero.

Under different ECG lead sources, the highest F1-scores in the most enhanced CAs classification tasks were achieved under $L_J$, with the exception of V1, V2, and V3. In Fig. 6, the ECG waveforms of a lead II heartbeat show that: (1) $L_R$ has better reconstruction power with the lowest MSE (0.14 %) and highest R (97.10 %); (2) ECG signals enhanced by $L_C$ may not present a clear reconstruction capability (MSE:35.05%; R:-2.72%) and resemble ECG signals. Such enhancements derived higher F1-scores than those obtained by $L_R$ (Table II), suggesting that the comprehension of machines can sometimes be beyond human intuition. (3) ECG signals enhanced by $L_J$ share the characteristics of those enhanced by $L_R$ and $L_C$ (MSE:5.40%; R:50.49%). Thus, $L_C$ may work as a constraint as both a penalty and incentive, restraining SRECG from overfitting and concentrating on crucial latent features.

### *D. PPRs in CAs classification with SRECG by $L_J$*

The previous F1-scores (Table II) demonstrated that, in most cases, SRECG by $L_J$ can increase the CAs classification accuracy of HMC (trained at 250 Hz) from low-resolution ECG signals (25 Hz). Therefore, another experiment with $L_J$ was conducted to further investigate the enhancement behavior of CAs classification accuracy, with PPR as a new metric for different ECG lead sources and CAs. The results in Fig. 7 show that the median PPR can reach up to overall 58%, yet variations across different leads and CAs exist. To visualize the median PPRs among the different leads and CAs, we labeled the median PPRs with color gradients based on their values. The colors from blue to green correspond to the median PPR values from high to low, and the black and red colors indicate anomalous median PPR values. The white color indicates the median PPR value that is not applicable (n/a) because its $C_P$, $C_N$, and $E_R$ are all zero.

From the color labeling in Fig. 7, it is noted that: (1) lead aVR, V4, V5, and V6 are shown to have great potential in the recovery of overall CAs classification accuracy (median PPR ranging from to 56-58%); (2) in contrast, lead V3 has the lowest median overall PPR:16%; (3) the recovery capability of different CAs is also observed to vary from lead to lead, for example, STE could be well recovered in lead I, II, III, aVR, V2, V3, V5, and V6 (median PPR ≥ 50%), but not in aVF, V1, and V4 (median PPR < 50%). However, there are six anomalies (median PPR < 0% or median PPR ≥ 100%) in Fig. 7, which in turn indicates that the F1-scores of HMCs with SRECG-enhanced ECG input are worse than those of $C_N$ (median PPR < 0%) or better than $C_P$ (median PPR ≥ 100%). This leads us to wonder if the enhanced signals may have unexpected predictive power changes for verifying the CAs.

### *E. Wilcoxon signed-rank test for four PPR anomalies in CAs classification with SRECG by $L_J$*

Conventionally, in statistics, a Wilcoxon signed-rank test is used to confirm that two sampled populations share significant differences if their p-value is less than 0.05. Therefore, the above six anomalies were subjected to the Wilcoxon signed-rank test in comparison to their $C_N$ or $C_P$, which are the 10-fold test F1-scores of low-resolution multiclass CAs classifiers trained at 25 Hz ECG signals (median PPR < 0%) or 250 Hz ECG signals (median PPR ≥ 100%). However, we found that none of the anomalies passed the criteria (p-value < 0.05); therefore, we cannot assert that there are significant differences compared with $C_N$ or $C_P$. Thus, we could not conclude that there were unexpected predictive power changes for verifying CAs in these six anomalies. Although the case of LBBB in lead V1 is still intriguing with a large increase (median PPR:131%), the main reason was no significant difference between $C_N$ and $C_P$ such that $C_P - C_N \approx 0$ which led to an enlarged value in PPR=$\frac{E_R - C_N}{C_P - C_N}$ according to its definition. Therefore, its increment of PPR is not valid.

## IV. Discussion

CE play an important role in smart healthcare to improve the quality of life. P/W devices incorporated with CE used for medical purposes have provided simple ways to monitor the health conditions of consumers in daily life, such as bracelets and wrist gadgets [68]. Several studies related to smart healthcare systems have utilized such devices to support visually impaired individuals and prevent falls in the elderly [69-71]. In addition, these devices can provide long-term physiological signal monitoring for clinical diagnosis and evaluation, such as electroencephalogram [72], ECG [73-75], and near-infrared spectroscopy [76].

Collectively, the combination of cloud-based DL algorithms with P/W devices is becoming a trend in smart healthcare systems. However, there are certain limits tied to the current capability of the hardware, such as battery life and computing power, for ECG long-term monitoring systems implemented in CE. With respect to the limitation on energy consumption, it is understood that more power is required when a higher frequency is used to record ECG signals [9-11]. Therefore, in this study, we attempted to alleviate this problem by recovering the predictive power of the HMC with a low sampling frequency input via the proposed DL-based SRECG framework, which is a novel SR-based method. The proposed SRECG is designed for P/W devices with the cloud-based workflow of a smart healthcare system (see Fig. 1).

For a conventional SR task, the loss function is typically only regression loss $L_R$, whereas our experiments indicated that enhancements by $L_R$ alone did not suffice to improve the CAs classification accuracies of the HMCs. As a result, it is viable to introduce the concept of teacher–student learning by including the CCE loss using the judge model's predictions, which are from a well-pretrained HMC for CAs classification [77]. Correspondingly, the experimental results showed that our joint loss $L_J$ obtained better performance on CAs classification (via F1-scores) in most of the ECG lead sources, with a few exceptions for leads V1, V2, and V3 (Table II). Moreover, the predictive power in the models of lead V5 trained at 250 Hz sampled ECG signals achieved up to 58% of their median overall PPR in the classifications of 25 Hz sampled ECG signals (see Fig. 7). It is interesting that the best reconstructed ECG signals (with the lowest MSEs and highest Rs) did not have the highest F1-scores when applying to HMCs for CAs classification. A possible reason for this is that the proposed model combining reconstruction and classification tasks pays more attention to reconstructing the features or signals that could help in ECG classification. In other words, the proposed approach is limited to processing fine-grained ECG signals that are relatively ineffective for classification tasks. Therefore, it could reach the highest model F1-scores without holding the lowest MSEs and the highest Rs. A similar phenomenon occurs in the field of automatic speech recognition (ASR) in a noisy environment [78-80]. They pointed out that removing background noise using speech enhancement models can improve signal quality, but it often degrades ASR performance. This is because the distortion caused by speech enhancement often deteriorates the ASR performance.

Clinically, DL-based ECG diagnosis of CAs has shown its feasibility and significance in improving diagnosis accuracy compared to that of general physicians and cardiologists [81, 82]. Different leads of ECG signals provide different ECG features for physicians and cardiologists to recognize specific types of arrhythmia [83-87]. Our experimental results in Fig. 7 show that lead V6 obtained the highest and standout median PPR of 98% in LBBB among all leads, which concurs with the clinically known LBBB diagnosis criteria by distinguishing QRS complex morphology at leads I, aVL, V1, V2, V5, and V6 [83]. In contrast, lead aVR also demonstrated the highest and distinguished median PPR of 81% in PVC, similar to lead V4 in STD (median PPR:70%) and lead V6 in AF (median PPR:67%). These observations suggest that the subtle ECG features hidden in leads aVR, V4, and V6 may serve as important judging criteria for DL models to distinguish PVC, STD, and AF, instead of human eyes.

Although the results demonstrated the capability to recover the predictive power of HMC trained at high-resolution ECG signals (250 Hz) with low-resolution input (25 Hz) by our SRECG, it becomes more challenging when trying to recover from even lower-resolution signals (< 25 Hz) to higher-resolution signals (> 250 Hz). Our method requires more validation on various types of datasets, for example, noisy datasets collected by embedded devices, to verify the prediction stability and effectiveness in practical situations. In the absence of a systematic evaluation approach and the lack of standardized datasets, it is difficult to address these limitations at the present stage. We consider these obstacles as part of our future study.

On the other hand, with the increasing concern about medical data privacy issues [88], it is necessary to propose encryption methods to make the data undistinguishable to satisfy the requirements of new privacy-preservation regulations, for example, GDPR [89]. In Fig. 6, we accidentally found that the ECG signals enhanced by the classification loss $L_C$ and joint loss $L_J$ can generate undistinguishable ECG waveforms while improving the predictive power for CAs classification (Table II), which may provide an alternative way to encrypt ECG data.

As machine learning and DL methods continue to advance and become user-friendly via P/W, the role of ECG will gradually open as a physiological information container. Subsequently, a DL model can be used at the right time to extract information from the container for ignored or unattended predictions such as gender and age [90]. Therefore, it is promising to expect ECG to provide more diagnostic power in many upcoming applications on P/W devices.

## V. Conclusion

To support automatic clinical diagnosis and long-term smart healthcare services on P/W devices while incorporated with CE, this work develops a novel DL-based SRECG as a cloud processor to enhance low-resolution ECG signals sampled from low frequency for CAs classification of HMC to achieve better accuracy. In future work, we will not only further investigate the effect of training SRECG with compound loss using different lead sources and their combinations but also explore different preprocessing methods for SRECG. Since DL-based diagnosis of ECG signals has great potential in deciphering other physiological conditions, it is worth applying SRECG for P/W devices in other diagnostic criteria of healthcare, such as mortality and heart failure, in addition to CAs.

## VI. References

[1] H. Zhu *et al.*, "Smart Healthcare in the Era of Internet-of-Things," *IEEE Consumer Electronics Magazine,* vol. 8, no. 5, pp. 26-30, 2019, doi: 10.1109/MCE.2019.2923929.

[2] S. P. Mohanty, U. Choppali, and E. Kougianos, "Everything you wanted to know about smart cities: The Internet of things is the backbone," *IEEE Consumer Electronics Magazine,* vol. 5, no. 3, pp. 60-70, 2016, doi: 10.1109/MCE.2016.2556879.

[3] A. S. Kibos, B. P. Knight, V. Essebag, S. B. Fishberger, M. Slevin, and I. C. Ţintoiu, *Cardiac Arrhythmias: From Basic Mechanism to State-of-the-Art Management*. Springer London, 2013.

[4] M. Shao, Z. Zhou, G. Bin, Y. Bai, and S. Wu, "A Wearable Electrocardiogram Telemonitoring System for Atrial Fibrillation Detection," *Sensors (Basel),* vol. 20, no. 3, Jan 22 2020, doi: 10.3390/s20030606.

[5] C. T. Lin *et al.*, "An intelligent telecardiology system using a wearable and wireless ECG to detect atrial fibrillation," *IEEE Trans Inf Technol Biomed,* vol. 14, no. 3, pp. 726-33, May 2010, doi: 10.1109/TITB.2010.2047401.

[6] S. Wattal, S. K. Spear, M. H. Imtiaz, and E. Sazonov, "A polypyrrole-coated textile electrode and connector for wearable ECG monitoring," in *2018 IEEE 15th International Conference on Wearable and Implantable Body Sensor Networks (BSN)*, 4-7 March 2018 2018, pp. 54-57, doi: 10.1109/BSN.2018.8329657.

[7] S. Hong *et al.*, "CardioLearn: A Cloud Deep Learning Service for Cardiac Disease Detection from Electrocardiogram," presented at the Companion Proceedings of the Web Conference 2020, Taipei, Taiwan, 2020. [Online]. Available: https://doi.org/10.1145/3366424.3383529.

[8] X. Wang, Q. Gui, B. Liu, Z. Jin, and Y. Chen, "Enabling Smart Personalized Healthcare: A Hybrid Mobile-Cloud Approach for ECG Telemonitoring," *IEEE Journal of Biomedical and Health Informatics,* vol. 18, no. 3, pp. 739-745, 2014, doi: 10.1109/JBHI.2013.2286157.

[9] J. Kim and C. Chu, "Analysis of energy consumption for wearable ECG devices," in *SENSORS, 2014 IEEE*, 2-5 Nov. 2014 2014, pp. 962-965, doi: 10.1109/ICSENS.2014.6985162.

[10] K. C. Liu, K. H. Hung, C. Y. Hsieh, H. Y. Huang, C. T. Chan, and Y. Tsao, "Deep Learning Based Signal Enhancement of Low-Resolution Accelerometer for Fall Detection Systems," *IEEE Transactions on Cognitive and Developmental Systems,* pp. 1-1, 2021, doi: 10.1109/TCDS.2021.3116228.

[11] X. Fafoutis, L. Marchegiani, A. Elsts, J. Pope, R. Piechocki, and I. Craddock, "Extending the battery lifetime of wearable sensors with embedded machine learning," in *2018 IEEE 4th World Forum on Internet of Things (WF-IoT)*, 5-8 Feb. 2018 2018, pp. 269-274, doi: 10.1109/WF-IoT.2018.8355116.

[12] N. Petrellis, I.-E. Kosmadakis, M. Vardakas, F. Gioulekas, M. Birbas, and A. Lalos, "Compressing and Filtering Medical Data in a Low Cost Health Monitoring System," in *Proceedings of the 21st Pan-Hellenic Conference on Informatics*, 2017, pp. 1-5.

[13] P. Maji, H. K. Mondal, A. P. Roy, S. Poddar, and S. P. Mohanty, "iKardo: An Intelligent ECG Device for Automatic Critical Beat Identification for Smart Healthcare," *IEEE Transactions on Consumer Electronics,* vol. 67, no. 4, pp. 235-243, 2021, doi: 10.1109/TCE.2021.3129316.

[14] T. Zhang, R. Ramakrishnan, and M. Livny, "BIRCH: an efficient data clustering method for very large databases," *SIGMOD Rec.,* vol. 25, no. 2, pp. 103–114, 1996, doi: 10.1145/235968.233324.

[15] N. V. Chawla, K. W. Bowyer, L. O. Hall, and W. P. Kegelmeyer, "SMOTE: synthetic minority over-sampling technique," *J. Artif. Int. Res.,* vol. 16, no. 1, pp. 321–357, 2002.

[16] G. Biagetti, P. Crippa, L. Falaschetti, S. Orcioni, and C. Turchetti, "Wireless surface electromyograph and electrocardiograph system on 802.15.4," *IEEE Transactions on Consumer Electronics,* vol. 62, no. 3, pp. 258-266, 2016, doi: 10.1109/TCE.2016.7613192.

[17] S. Raj, "An Efficient IoT-Based Platform for Remote Real-Time Cardiac Activity Monitoring," *IEEE Transactions on Consumer Electronics,* vol. 66, no. 2, pp. 106-114, 2020, doi: 10.1109/TCE.2020.2981511.

[18] Apple. "Taking an ECG with the ECG app on Apple Watch Series 4." https://support.apple.com/hr-hr/HT208955 (accessed 4/20, 2019).

[19] L. S. Brunner, S. C. O. C. Smeltzer, B. G. Bare, J. L. Hinkle, and K. H. Cheever, *Brunner & Suddarth's Textbook of Medical-surgical Nursing* (no. 第 1 卷). Wolters Kluwer Health/Lippincott Williams & Wilkins, 2010.

[20] T. M. Chen, C. H. Huang, E. S. C. Shih, Y. F. Hu, and M. J. Hwang, "Detection and Classification of Cardiac Arrhythmias by a Challenge-Best Deep Learning Neural Network Model," *iScience,* vol. 23, no. 3, p. 100886, Feb 4 2020, doi: 10.1016/j.isci.2020.100886.

[21] S. Mahdiani, V. Jeyhani, M. Peltokangas, and A. Vehkaoja, "Is 50 Hz high enough ECG sampling frequency for accurate HRV analysis?," *Conf Proc IEEE Eng Med Biol Soc,* vol. 2015, pp. 5948-51, 2015, doi: 10.1109/EMBC.2015.7319746.

[22] O. Kwon *et al.*, "Electrocardiogram Sampling Frequency Range Acceptable for Heart Rate Variability Analysis," *Healthc Inform Res,* vol. 24, no. 3, pp. 198-206, Jul 2018, doi: 10.4258/hir.2018.24.3.198.

[23] K. A. Sidek and I. Khalil, "Enhancement of low sampling frequency recordings for ECG biometric matching using interpolation," *Computer Methods and Programs in Biomedicine,* vol. 109, no. 1,

pp. 13-25, 2013/01/01/ 2013, doi: https://doi.org/10.1016/j.cmpb.2012.08.015.

[24] S. M. Mathews, C. Kambhamettu, and K. E. Barner, "A novel application of deep learning for single-lead ECG classification," (in eng), *Comput Biol Med,* vol. 99, pp. 53-62, Aug 1 2018, doi: 10.1016/j.compbiomed.2018.05.013.

[25] Y. L. Coelho, F. d. A. S. d. Santos, A. Frizera-Neto, and T. F. Bastos-Filho, "A Lightweight Framework for Human Activity Recognition on Wearable Devices," *IEEE Sensors Journal,* vol. 21, no. 21, pp. 24471-24481, 2021, doi: 10.1109/JSEN.2021.3113908.

[26] Z. Long, T. Wang, C. You, Z. Yang, K. Wang, and J. Liu, "Terahertz image super-resolution based on a deep convolutional neural network," *Appl Opt,* vol. 58, no. 10, pp. 2731-2735, Apr 1 2019, doi: 10.1364/AO.58.002731.

[27] M. Kwon, S. Han, K. Kim, and S. C. Jun, "Super-Resolution for Improving EEG Spatial Resolution using Deep Convolutional Neural Network-Feasibility Study," *Sensors (Basel),* vol. 19, no. 23, Dec 3 2019, doi: 10.3390/s19235317.

[28] K. Umehara, J. Ota, and T. Ishida, "Application of Super-Resolution Convolutional Neural Network for Enhancing Image Resolution in Chest CT," *J Digit Imaging,* vol. 31, no. 4, pp. 441-450, Aug 2018, doi: 10.1007/s10278-017-0033-z.

[29] D. Dai, Y. Wang, Y. Chen, and L. Van Gool, "Is Image Super-resolution Helpful for Other Vision Tasks?," p. arXiv:1509.07009. [Online]. Available: https://ui.adsabs.harvard.edu/abs/2015arXiv150907009D

[30] M. Haris, G. Shakhnarovich, and N. Ukita, "Task-Driven Super Resolution: Object Detection in Low-resolution Images," p. arXiv:1803.11316. [Online]. Available: https://ui.adsabs.harvard.edu/abs/2018arXiv180311316H

[31] M. S. M. Sajjadi, B. Schölkopf, and M. Hirsch, "EnhanceNet: Single Image Super-Resolution Through Automated Texture Synthesis," p. arXiv:1612.07919. [Online]. Available: https://ui.adsabs.harvard.edu/abs/2016arXiv161207919S

[32] Y. Bai, Y. Zhang, M. Ding, and B. Ghanem, "SOD-MTGAN: Small Object Detection via Multi-Task Generative Adversarial Network," in *Computer Vision – ECCV 2018*, Cham, V. Ferrari, M. Hebert, C. Sminchisescu, and Y. Weiss, Eds., 2018// 2018: Springer International Publishing, pp. 210-226.

[33] W.-S. Lai, J.-B. Huang, N. Ahuja, and M.-H. Yang, "Deep Laplacian Pyramid Networks for Fast and Accurate Super-Resolution," p. arXiv:1704.03915. [Online]. Available: https://ui.adsabs.harvard.edu/abs/2017arXiv170403915L

[34] C. Dong, C. C. Loy, K. He, and X. Tang, "Learning a Deep Convolutional Network for Image Super-Resolution," Cham, 2014: Springer International Publishing, in Computer Vision – ECCV 2014, pp. 184-199.

[35] J. Kim, J. K. Lee, and K. M. Lee, "Accurate Image Super-Resolution Using Very Deep Convolutional Networks," p. arXiv:1511.04587. [Online]. Available: https://ui.adsabs.harvard.edu/abs/2015arXiv151104587K

[36] N. Ahn, B. Kang, and K.-A. Sohn, "Fast, Accurate, and Lightweight Super-Resolution with Cascading Residual Network," p. arXiv:1803.08664. [Online]. Available: https://ui.adsabs.harvard.edu/abs/2018arXiv180308664A

[37] J. Johnson, A. Alahi, and L. Fei-Fei, "Perceptual Losses for Real-Time Style Transfer and Super-Resolution," p. arXiv:1603.08155. [Online]. Available: https://ui.adsabs.harvard.edu/abs/2016arXiv160308155J

[38] A. Bulat and G. Tzimiropoulos, "Super-FAN: Integrated facial landmark localization and super-resolution of real-world low resolution faces in arbitrary poses with GANs," p. arXiv:1712.02765. [Online]. Available: https://ui.adsabs.harvard.edu/abs/2017arXiv171202765B

[39] C. Ledig *et al.*, "Photo-Realistic Single Image Super-Resolution Using a Generative Adversarial Network," p. arXiv:1609.04802. [Online]. Available: https://ui.adsabs.harvard.edu/abs/2016arXiv160904802L

[40] B. Lim, S. Son, H. Kim, S. Nah, and K. M. Lee, "Enhanced Deep Residual Networks for Single Image Super-Resolution," p. arXiv:1707.02921. [Online]. Available: https://ui.adsabs.harvard.edu/abs/2017arXiv170702921L

[41] Y. Wang, F. Perazzi, B. McWilliams, A. Sorkine-Hornung, O. Sorkine-Hornung, and C. Schroers, "A Fully Progressive Approach to Single-Image Super-Resolution," p. arXiv:1804.02900. [Online]. Available: https://ui.adsabs.harvard.edu/abs/2018arXiv180402900W

[42] Y. Zhang, K. Li, K. Li, L. Wang, B. Zhong, and Y. Fu, "Image Super-Resolution Using Very Deep Residual Channel Attention Networks," p. arXiv:1807.02758. [Online]. Available: https://ui.adsabs.harvard.edu/abs/2018arXiv180702758Z

[43] T. Dai, J. Cai, Y. Zhang, S. Xia, and L. Zhang, "Second-Order Attention Network for Single Image Super-Resolution," in *2019 IEEE/CVF Conference on Computer Vision and Pattern Recognition (CVPR)*, 15-20 June 2019 2019, pp. 11057-11066, doi: 10.1109/CVPR.2019.01132.

[44] G. Hinton *et al.*, "Deep Neural Networks for Acoustic Modeling in Speech Recognition: The Shared Views of Four Research Groups," *IEEE Signal Processing Magazine,* vol. 29, no. 6, pp. 82-97, 2012, doi: 10.1109/MSP.2012.2205597.

[45] D. Jurafsky and J. H. Martin, "Speech and language processing," (in English), 2014. [Online]. Available: http://www.dawsonera.com/depp/reader/protected/external/AbstractView/S9781292037936.

[46] X. Lu, Y. Tsao, S. Matsuda, and C. Hori, "Speech enhancement based on deep denoising autoencoder," in *INTERSPEECH*, 2013.

[47] A. van den Oord *et al.*, "WaveNet: A Generative Model for Raw Audio," p. arXiv:1609.03499. [Online]. Available: https://ui.adsabs.harvard.edu/abs/2016arXiv160903499V

[48] J.-P. Briot, G. Hadjeres, and F.-D. Pachet, "Deep Learning Techniques for Music Generation -- A Survey," p. arXiv:1709.01620. [Online]. Available: https://ui.adsabs.harvard.edu/abs/2017arXiv170901620B

[49] M. A. Acevedo, C. J. Corrada-Bravo, H. Corrada-Bravo, L. J. Villanueva-Rivera, and T. M. Aide, "Automated classification of bird and amphibian calls using machine learning: A comparison of methods," *Ecological Informatics,* vol. 4, no. 4, pp. 206-214, 2009, doi: 10.1016/j.ecoinf.2009.06.005.

[50] S. Mehri *et al.*, "SampleRNN: An Unconditional End-to-End Neural Audio Generation Model," p. arXiv:1612.07837. [Online]. Available: https://ui.adsabs.harvard.edu/abs/2016arXiv161207837M

[51] C. L. Liu, S. W. Fu, Y. J. Li, J. W. Huang, H. M. Wang, and Y. Tsao, "Multichannel Speech Enhancement by Raw Waveform-Mapping Using Fully Convolutional Networks," *IEEE/ACM Transactions on Audio, Speech, and Language Processing,* vol. 28, pp. 1888-1900, 2020, doi: 10.1109/TASLP.2020.2976193.

[52] S. W. Fu, T. W. Wang, Y. Tsao, X. Lu, and H. Kawai, "End-to-End Waveform Utterance Enhancement for Direct Evaluation Metrics Optimization by Fully Convolutional Neural Networks," *IEEE/ACM Transactions on Audio, Speech, and Language Processing,* vol. 26, no. 9, pp. 1570-1584, 2018, doi: 10.1109/TASLP.2018.2821903.

[53] S. Birnbaum, V. Kuleshov, Z. Enam, P. W. Koh, and S. Ermon, "Temporal FiLM: Capturing Long-Range Sequence Dependencies with Feature-Wise Modulations," p. arXiv:1909.06628. [Online]. Available: https://ui.adsabs.harvard.edu/abs/2019arXiv190906628B

[54] F. Liu *et al.*, "An Open Access Database for Evaluating the Algorithms of Electrocardiogram Rhythm and Morphology Abnormality Detection," *Journal of Medical Imaging and Health Informatics,* vol. 8, no. 7, pp. 1368-1373, // 2018, doi: 10.1166/jmihi.2018.2442.

[55] A. L. Goldberger *et al.*, "PhysioBank, PhysioToolkit, and PhysioNet: components of a new research resource for complex physiologic signals," *Circulation,* vol. 101, no. 23, pp. e215-e220, 2000.

[56] G. B. Moody and R. G. Mark, "The impact of the MIT-BIH arrhythmia database," *IEEE Engineering in Medicine and Biology Magazine,* vol. 20, no. 3, pp. 45-50, 2001.

[57] Z. Yang, D. Yang, C. Dyer, X. He, A. J. Smola, and E. H. Hovy, "Hierarchical Attention Networks for Document Classification," in *HLT-NAACL*, 2016.

[58] M. Schuster and K. K. Paliwal, "Bidirectional recurrent neural networks," (in English), *Ieee T Signal Proces,* vol. 45, no. 11, pp. 2673-2681, Nov 1997, doi: Doi 10.1109/78.650093.

[59] J. Hearty, *Advanced Machine Learning with Python*. Packt Publishing, 2016.

[60] S. Ioffe and C. Szegedy, "Batch Normalization: Accelerating Deep Network Training by Reducing Internal Covariate Shift," presented

at the Proceedings of the 32nd International Conference on Machine Learning, Proceedings of Machine Learning Research, 2015. [Online]. Available: https://proceedings.mlr.press/v37/ioffe15.html.

[61] B. Xu, N. Wang, T. Chen, and M. Li, "Empirical Evaluation of Rectified Activations in Convolutional Network," p. arXiv:1505.00853. [Online]. Available: https://ui.adsabs.harvard.edu/abs/2015arXiv150500853X

[62] M. Abadi *et al.*, "TensorFlow: Large-Scale Machine Learning on Heterogeneous Distributed Systems," *ArXiv e-prints*, vol. 1603. [Online]. Available: http://adsabs.harvard.edu/abs/2016arXiv160304467A

[63] P. W. D. Charles, "Project Title," *GitHub repository,* vol. https://github.com/charlespwd/project-title, 2013.

[64] F. Pedregosa *et al.*, "Scikit-learn: Machine Learning in Python," *J. Mach. Learn. Res.,* vol. 12, no. null, pp. 2825–2830, 2011.

[65] D. P. Kingma and J. Ba, "Adam: A Method for Stochastic Optimization," *ArXiv e-prints*, vol. 1412. [Online]. Available: http://adsabs.harvard.edu/abs/2014arXiv1412.6980K

[66] C. Ledig *et al.*, "Photo-Realistic Single Image Super-Resolution Using a Generative Adversarial Network," in *2017 IEEE Conference on Computer Vision and Pattern Recognition (CVPR)*, 21-26 July 2017 2017, pp. 105-114, doi: 10.1109/CVPR.2017.19.

[67] K. He, X. Zhang, S. Ren, and J. Sun, "Deep Residual Learning for Image Recognition," in *2016 IEEE Conference on Computer Vision and Pattern Recognition (CVPR)*, 27-30 June 2016 2016, pp. 770-778, doi: 10.1109/CVPR.2016.90.

[68] H. Thapliyal, V. Khalus, and C. Labrado, "Stress Detection and Management: A Survey of Wearable Smart Health Devices," *IEEE Consumer Electronics Magazine,* vol. 6, no. 4, pp. 64-69, 2017, doi: 10.1109/MCE.2017.2715578.

[69] S. H. Chae, M. C. Kang, J. Y. Sun, B. S. Kim, and S. J. Ko, "Collision detection method using image segmentation for the visually impaired," *IEEE Transactions on Consumer Electronics,* vol. 63, no. 4, pp. 392-400, 2017, doi: 10.1109/TCE.2017.015101.

[70] C. W. Lee, P. Chondro, S. J. Ruan, O. Christen, and E. Naroska, "Improving Mobility for the Visually Impaired: A Wearable Indoor Positioning System Based on Visual Markers," *IEEE Consumer Electronics Magazine,* vol. 7, no. 3, pp. 12-20, 2018, doi: 10.1109/MCE.2018.2797741.

[71] J. Wang, Z. Zhang, B. Li, S. Lee, and R. S. Sherratt, "An enhanced fall detection system for elderly person monitoring using consumer home networks," *IEEE Transactions on Consumer Electronics,* vol. 60, no. 1, pp. 23-29, 2014, doi: 10.1109/TCE.2014.6780921.

[72] M. A. Sayeed, S. P. Mohanty, E. Kougianos, and H. P. Zaveri, "Neuro-Detect: A Machine Learning-Based Fast and Accurate Seizure Detection System in the IoMT," *IEEE Transactions on Consumer Electronics,* vol. 65, no. 3, pp. 359-368, 2019, doi: 10.1109/TCE.2019.2917895.

[73] N. Dey, A. S. Ashour, F. Shi, S. J. Fong, and R. S. Sherratt, "Developing residential wireless sensor networks for ECG healthcare monitoring," *IEEE Transactions on Consumer Electronics,* vol. 63, no. 4, pp. 442-449, 2017, doi: 10.1109/TCE.2017.015063.

[74] S. Y. Lee, P. W. Huang, M. C. Liang, J. H. Hong, and J. Y. Chen, "Development of an Arrhythmia Monitoring System and Human Study," *IEEE Transactions on Consumer Electronics,* vol. 64, no. 4, pp. 442-451, 2018, doi: 10.1109/TCE.2018.2875799.

[75] S. Raj and K. C. Ray, "A Personalized Point-of-Care Platform for Real-Time ECG Monitoring," *IEEE Transactions on Consumer Electronics,* vol. 64, no. 4, pp. 452-460, 2018, doi: 10.1109/TCE.2018.2877481.

[76] A. M. Joshi, P. Jain, and S. P. Mohanty, "iGLU 3.0: A Secure Noninvasive Glucometer and Automatic Insulin Delivery System in IoMT," *IEEE Transactions on Consumer Electronics,* vol. 68, no. 1, pp. 14-22, 2022, doi: 10.1109/TCE.2022.3145055.

[77] J. Li, M. L. Seltzer, X. Wang, R. Zhao, and Y. Gong, "Large-Scale Domain Adaptation via Teacher-Student Learning," p. arXiv:1708.05466. [Online]. Available: https://ui.adsabs.harvard.edu/abs/2017arXiv170805466L

[78] S. O. Sadjadi and J. H. Hansen, "Assessment of single-channel speech enhancement techniques for speaker identification under mismatched conditions," in *Eleventh Annual Conference of the International Speech Communication Association*, 2010.

[79] M. Fujimoto and H. Kawai, "One-Pass Single-Channel Noisy Speech Recognition Using a Combination of Noisy and Enhanced Features," in *INTERSPEECH*, 2019, pp. 486-490.

[80] H. Sato, T. Ochiai, M. Delcroix, K. Kinoshita, N. Kamo, and T. Moriya, "Learning to Enhance or Not: Neural Network-Based Switching of Enhanced and Observed Signals for Overlapping Speech Recognition," in *ICASSP 2022-2022 IEEE International Conference on Acoustics, Speech and Signal Processing (ICASSP)*, 2022: IEEE, pp. 6287-6291.

[81] A. Y. Hannun *et al.*, "Cardiologist-level arrhythmia detection and classification in ambulatory electrocardiograms using a deep neural network," *Nature medicine,* vol. 25, no. 1, p. 65, 2019.

[82] A. Shiyovich, A. Wolak, L. Yacobovich, A. Grosbard, and A. Katz, "Accuracy of diagnosing atrial flutter and atrial fibrillation from a surface electrocardiogram by hospital physicians: analysis of data from internal medicine departments," *The American journal of the medical sciences,* vol. 340, no. 4, pp. 271-275, 2010.

[83] A. B. de Luna, *Clinical Electrocardiography, Enhanced Edition: A Textbook*. Wiley, 2012.

[84] K. Wesley, *Huszar's ECG and 12-Lead Interpretation - E-Book*. Elsevier Health Sciences, 2016.

[85] Y. Kobayashi, "Idiopathic Ventricular Premature Contraction and Ventricular Tachycardia: Distribution of the Origin, Diagnostic Algorithm, and Catheter Ablation," *Journal of Nippon Medical School,* vol. 85, no. 2, pp. 87-94, 2018.

[86] T. Garcia and G. Miller, *Arrhythmia Recognition: The Art of Interpretation*. Jones & Bartlett Learning, 2004.

[87] E. B. Hanna and D. L. Glancy, "ST-segment depression and T-wave inversion: classification, differential diagnosis, and caveats," *Cleveland Clinic journal of medicine,* vol. 78, no. 6, p. 404, 2011.

[88] Q. Yang, Y. Liu, T. Chen, and Y. Tong, "Federated Machine Learning: Concept and Applications," p. arXiv:1902.04885. [Online]. Available: https://ui.adsabs.harvard.edu/abs/2019arXiv190204885Y

[89] F. Fatehi, F. Hassandoust, R. K. L. Ko, and S. Akhlaghpour, "General Data Protection Regulation (GDPR) in Healthcare: Hot Topics and Research Fronts," *Stud Health Technol Inform,* vol. 270, pp. 1118-1122, Jun 16 2020, doi: 10.3233/SHTI200336.

[90] Z. I. Attia *et al.*, "Age and Sex Estimation Using Artificial Intelligence From Standard 12-Lead ECGs," *Circ Arrhythm Electrophysiol,* vol. 12, no. 9, p. e007284, Sep 2019, doi: 10.1161/CIRCEP.119.007284.